\documentclass[12pt]{article}  

\usepackage{epsf}  
\usepackage{amsfonts}  
\usepackage{amssymb}  
 
\catcode `@=11
\@addtoreset{equation}{section} 
\catcode `@=12

\newcommand{\f}{\frac}
\newcommand{\half}{\frac12}

\newcommand{\beq}{\begin{eqnarray}}% can be used as {equation} or {eqnarray}
\newcommand{\eeq}{\end{eqnarray}}

\def\sector#1#2{\ {\scriptstyle #1}\hskip 1mm  
\mathop{\square}\limits_{\lower 1mm\hbox{$\scriptstyle#2$}}\hskip 1mm}  
  
\def\tsector#1#2{\ {\scriptstyle #1}\hskip 1mm  
\mathop{\opensquare}\limits_{\lower 1mm\hbox{$\scriptstyle#2$}}^\sim\hskip 1mm}  

\def\){\right)}
\def\({\left( }
\def\]{\right] }
\def\[{\left[ }

\def\IC{{\relax\hbox{$\inbar\kern-.3em{\rm C}$}}}  
\def\IZ{{\relax\hbox{$\inbar\kern-.3em{\rm Z}$}}}

\def\al{\alpha}

\def\be{\begin{equation}}
\def\ee{\end{equation}}
\def\bea{\begin{eqnarray}}
\def\eea{\end{eqnarray}}

\newcommand{\rem}[1]{}

\def\half{{1\over 2}}

\def\ltap{\ \raise.3ex\hbox{$<$\kern-.75em\lower1ex\hbox{$\sim$}}\ }
\def\gtap{\ \raise.3ex\hbox{$>$\kern-.75em\lower1ex\hbox{$\sim$}}\ }

\newcommand{\ba}{\begin{eqnarray}}
\newcommand{\ea}{\end{eqnarray}}
\newcommand{\no}{\nonumber \\}

\def\bea{\begin{eqnarray}}
\def\eea{\end{eqnarray}}

\def\half{\frac{1}{2}}

\def\Tr{{\rm Tr}}

\begin{document}

\title{Tachyon mass, c-function and \\Counting localized degrees of freedom.}
\author{ Sang-Jin Sin  \\ \\
 \small \sl 
Department of Physics, Hanyang University, Seoul, 133-791, 
Korea\\ \small {\tt sjs@hepth.hanyang.ac.kr} 
 \\ 
} \maketitle
\begin{abstract} 
We discuss the localized tachyon condensation in the 
non-supersymmetric orbifold theories by taking the cosmological constant 
as the measure of degrees of freedom (d.o.f). We first show asymptotic 
density of state is not a proper quantity to count the 'localized' d.o.f. 
We then show that localized  d.o.f lead us a c-function given by the 
lightest tachyon mass, which turns out to be the same as the tachyon potential 
recently suggested by Dabholkar and Vafa.  
We also argue that delocalized d.o.f also encode information on the process of
 localized tachyon condensation in the g-function, 
based on the fact that the global geometry of the orbifolds 
is completely determined by the local geometry around the fixed points. 
For type II, both c- and g-function respect the stability of the 
supersymmetric models and both allow all the process suggested by 
Adams, Polchinski and Silverstein. 
\end{abstract}

\newpage 
\section{Introduction and Summary}
Recently there has been some interests on tachyon condensation in closed
string theories \cite{Russo:2001tf,aps,Vafa:2001ra,Michishta:2001ph,David:2001vm}. 
Closed string tachyons indicate the decay of the spacetime
itself and full dynamical understanding is still lacking. 
To simplify the problem and to utilize the experience from the open string
case \cite{sen}  Adams, Polchinski, and Silverstein (APS) \cite{aps}
studied the cases where the tachyons are localized at the tip of the orbifolds. 
They argued that the tachyon  condensation causes the  orbifolds to decay and
the tip of the orbifolds smooth out. They supported this conjecture by 
 D-brane probes  in sub-stringy regime and by general relativity analysis beyond string scale.
According to the analysis of APS, the tachyon condensation induces 
cascade of phase transitions until all the tachyons disappear and supersymmetry restored. 
So each step of cascade involve a RG flow along which IR is less tachyonic, less singular, and more 
supersymmetric than UV.  
 
Having acquired the general tendency of localized tachyon condensation from the analysis of APS, 
it is most desired to have a $c$-function \cite{Zamolodchikov:gt}
which decreases along the possible RG-flows and summarize the possible decay modes. In this context, 
Dabholkar and Vafa (DV) \cite{dabholkar},  utilizing the 
worldsheet $N=2$ supersymmetry, have proposed a closed string 
tachyon action describing the real time tachyon dynamics, where  generalized 
c-function \cite{Cecotti:1991me} was adopted as the tachyon potential.
In the same context, Harvey, Kutasov, Martinec and Moore (HKMM) \cite{hkmm} suggested another 
criteria based on counting the asymptotic density of states(ADOS), 
which suggested that not all examples of APS is consistent with 
their analysis.  In Ref.\cite{ns1}, we proposed a c-function of RG flow that is analogous but 
different from that suggested in \cite{hkmm}. The proposal in 
\cite{dabholkar,ns1} allows all of the process of APS, while 
that in \cite{hkmm} does not. Since   different proposals, which 
do not agree fully, are suggested, it is an urgent matter to 
resolve this issue. 

In this paper, we discuss this issue by counting the degrees of 
freedom (DOF)
from slightly different point of view. 
First we will show that, for the problem at hand, the cosmological constant(CC)  is a good measure 
of degree of freedom even in the presence of the spacetime fermions. 
We will then show that  while the cosmological constant  and the asymptotic density of states 
(ADOS)  yield the same central charge for the total DOF, 
they give different results for the localized DOF. 
After showing that the latter(ADOS) yields the bulk degree of freedom smaller than the 
localized degree of freedom, which is counter intuitive, we will suggest that  we have to use the 
former(CC).

We conclude that what measure the localized 
degree of freedom is the minimal twisted tachyon mass, which turns out to 
be equal to a c-function's value at conformal points. 
The result  does not depend on whether we use total partition function or
its bosonic part. 
 
The minimal tachyon mass can be shown to be equal to the 
deficit angle of the orbifold geometry. It is also equal to 
the tachyon potential suggested by Dabholkar and Vafa \cite{dabholkar}.  
The proposal that the absolute value of minimal tachyon mass as the 
c-function of the RG flow means that the theory becomes less tachyonic along 
the phase transition. It is certainly a generalized $c$-theorem,
proof of which, however, is not an issue here.

Although we mainly discuss the localized degree of freedom, 
the bulk degree of freedom also encodes some information of  
the orbifold geometry, because local geometry around the tip completely fix the global geometry
for the orbifolds. It turns out that considering the delocalized 
degree of freedom leads us  
g-function; $g_{cl}$ for type 0 as given in \cite{hkmm} and  
$g^{II}_{cl}$ as given in \cite{ns1}. 
So emerging picture is following: {\it localized DOF gives 
a c-function that is given by  minimal twisted  tachyon mass, 
while the delocalized DOF leads us to g-theorems. }

Our analysis seems to support the picture that the cosmological constant itself, 
although divergent due to the tachyons, can play the role of the tachyon potential in 
closed string theories. This is in a close analogy with the open 
string phenomena where world-sheet (sphere) partition function is 
the spacetime action. In the presence of the worldsheet 
supersymmetry, there is no potential coming from the sphere 
amplitude for the closed string tachyon \cite{tseytlin}, therefore 
 the 1-loop result should be the leading order tachyon potential. This picture predicts that  
IR is more supersymmetric,  less tachyonic and less singular than 
UV, which is  precisely what the analysis of Adams, Polchinski and 
Silverstein suggests.  We end the paper by possible future 
problems.   
   
\section{Cosmological constant v.s Asymptotic density of states}
We first consider  cosmological constant, namely, the  string 
vacuum amplitude on torus, 
\be
Z=\int_F \frac{d^2\tau}{\tau_2^2} Z(\tau), 
\ee
where $F$ represent  a fundamental domains of torus moduli, 
$|\tau|>1, |\tau_1|<\half, \tau=\tau_1+i\tau_2$. Notice that our 
fundamental domain does not contain the  $\tau_2 \to 0$ region.  
 $ Z(\tau)$ is the torus partition function. 

On the other hand, the general partition function can be written 
as follows: 
\be
Z(\tau)=\Tr \exp (-\tau_2 \pi \alpha' M^2 +2\pi i\tau_1(L_0-{\bar 
L}_0)),\label{ztau} 
\ee
where $M^2$ is the mass operator.  Therefore in the presence of 
tachyons, the dominant divergent part comes from the 
$\tau_2\to\infty$ region. In type 0 case, the bulk tachyon ( 
tachyon in untwisted sector) gives the dominant contribution.  In type II theory, 
it is projected out so the dominant contribution comes from the lightest twisted tachyon. 
Although the integral diverges, we are interested in the 
degree of divergence, $c_{eff}$, which  precisely is the information of degree of 
freedom. 
To do this, we can simply approximate the integral by 
the value of $Z(\tau_2)$ at $\tau_2=\infty, \tau_1=0$ multiplied 
by the area of fundamental domain, $\int_F 
d^2\tau/\tau_2^2=\frac{\pi}{3}$. One can do slightly better by 
averaging over $\tau_1$; 
\be
\int_F \frac{d^2\tau}{\tau_2^2} Z(\tau) = \frac{\pi}{3} 
\lim_{\tau_2\to\infty}\int_{-\half}^\half  d\tau_1 Z(\tau).
\ee
 This is nothing but a version of the  theorem  \cite{Kutasov:1990sv}
  by Kutasov and Seiberg in the presence of  tachyon.  
  For practical calculation, we can simply set 
$\tau_1=0$  with  the level matching condition  properly 
imposed. See Eq.(\ref{ztau}). So, if there are 
tachyons,  the cosmological constant can be simply related to the low temperature 
limit\footnote{$\tau_2$ is the inverse temperature $\beta$} of the partition 
function;
\be
Z = \lim_{\tau_2 \to \infty} \frac{\pi}{3}Z(\tau_2=\infty) = \lim_{\tau_2 
\to \infty} \exp(\pi\alpha'|M^2_{min}|\tau_2). 
\ee
Here, $M^2_{min}$ is minimal mass  of the twisted tachyons and we 
neglected power  of $\tau_2^{-3}$ as well as constant factors in 
front of the exponential divergence. 

In a general conformal field theory, the partition function in the 
low temperature limit gives the central charge by
\be
Z(\tau_2 \to \infty,\tau_1=0) \sim g {q}^{-\ {c_{eff}}/{12}}, 
\label{c} 
\ee  
where $ {q}=\exp{(-2\pi\tau_2)}$, $g$ is a co-efficient of the 
leading term and
\be 
{c_{\rm eff}}:=c - 24 \Delta_{min} 
\ee 
with  $\Delta_{min}={\rm min} \half(\Delta+{\bar \Delta})$. From 
Eqs.(\ref{ztau}) and (\ref{c}), the lightest tachyon mass  
naturally gives an effective central charge 
\be
{c_{eff}}=  6|{\rm min}\; \alpha ' M^2|. \label{d} 
\ee 
Therefore the precise definition of the  quantity we are interested in 
can be given as 
\be
c_{eff}:=\frac{6}{\pi} \lim_{\tau_2 \to \infty} \frac{1}{\tau_2}\log Z(\tau) 
.
\ee 
Since $c_{eff}$ is the measure of 
degree of freedom, so is  the cosmological constant.
On the other hand, there is well known measure 
of degrees of freedom which is a high energy density of states. It  
comes from the high temperature behaviour of the partition function;
\be
Z(\tau_2 \to 0) \sim {\tilde q}^{-\ {c_{eff}}/{12}}, 
\label{ac} 
\ee  
where $ {\tilde q}=\exp{(-2\pi/\tau_2)}$. 
When we deal with modular invariant (total) partition 
functions, two $c_{eff}$'s appearing in Eq.(\ref{c}) and  Eq.(\ref{ac}) are  the 
same, of course. However, for the localized degree of freedom states, 
two are different as we will show shortly.

In our case, $Z(\tau)$ is the partition function of the  CFT 
representing  lightcone string theory in 10 dimension. If the bulk 
tachyon is present, Eq.(\ref{d})  correctly identifies the 
transverse dimension $8$ out of central charge; $c=c_{eff}= 
6\times |-2|= 8\times 3/2$ from the tachyon spectrum. \footnote{If we consider the 
conformal field theory on  the orbifold  ${\bf C}^1/{\bf Z}_N$ 
rather than the full string partition function in 10 dimension, 
its partition function is above  $Z_(\tau)$ without $|\eta|^{-18} 
\sim q^{-9/12 }$. So the central charge 3 could have been  simply 
read off from the bulk tachyon. But there is no meaning of "mass 
spectrum", so is the word "tachyon". This is partly our motive to 
deal with sting partition function rather than the relevant 
orbifold CFT itself.} 
 
So far, we have seen that, in the presence of tachyons,  the cosmological constant or 
the low temperature limit of the partition function can a good 
measure of degree of freedom. However, if there are 
spacetime fermions, it is more subtle. 
For example, in the supersymmetric theory, the GSO projected partition function 
$Z=Z_B-Z_F$ is identically zero. Then when it correctly count the degree of freedom?
Looking at the eq. (\ref{c}) the answer is obvious: {\it As far as the 
dominant piece, $q^{-c_{eff}/12}$, is not cancelled, 
it is a good measure of degrees of freedom.} 

\section{Counting localized degree of freedom}

Now we discuss on the splitting the degree of freedoms into localized 
and delocalized ones. In ref. \cite{hkmm},
it is suggested that  localized degree of freedom is coming from 
the partition function of the twisted fields, and the delocalized (or bulk) degree of freedom is 
coming from the untwisted part;
\be
Z(\tau)=Z_{un}+Z_{tw}. 
\ee
Since the twisted field is fixed at the tip of the orbifolds,  
this is very reasonable suggestion. They proceeded to count the 
localized degree of freedom by looking at the ADOS coming from the high 
temperature behaviour of the twisted partition function, $Z_{tw}(\tau_2\to 0)$. 

\subsection{type 0: with bulk tachyon}
If we look at the high temperature limit, we get 
the result of HKMM \cite{hkmm};
\be 
Z_{tw}(\tau_2 \to 0) \sim g_{cl}{\tilde q}^{-1}, 
\ee
where $-1$ comes from the bulk tachyon spectrum $\half\alpha'M^2=-1$. 
It gives us $c=12$ suggesting that it is measuring {\it bulk} degrees of 
freedom, while we want to measure localized ones when we use $Z_{tw}$.
Notice that while $Z_{tw}(\tau_2)$ does not contain any 
bulk tachyon spectrum, the bulk tachyon spectrum is introduced by 
the modular transformation. Now, if we look at the 
delocalized degree of freedom, its high temperature limit gives 
\be 
Z_{un}(\tau_2 \to 0) \sim {\tilde q}^{\half\alpha'M_{min}^2}, 
\ee  
where $\half \alpha'M_{min}^2$ is the minimal mass of the twisted 
sector. 
Since  
\be 
\left| \half \alpha'M_{min}^2\right| =1-\frac{1}{N}<1 ,
\ee
for $C^1/Z_N$ model,  the localized degree of freedom  
is bigger than the bulk degree of freedom (DOF), which is counter-intuitive. 
Bulk and local DOF are interchanged. 
The reason of bulk-local interchange is obviously due to the 
modular transformation.  So, when we use modular 
non-invarinat object as a indicator of physical quantity, we need 
care.

On the other hand, if we use the cosmological constant, there is no modular 
transformation involved and we get intuitive results;
\be
Z_{tw}(\tau_2 \to \infty) \sim {  q}^{\half\alpha'M_{min}^2}, 
\ee
and 
\be 
Z_{un}(\tau_2 \to \infty) \sim g_{cl}{  q}^{-1}.  
\ee
Using the cosmological constant as the measure of degree of 
freedom is also consistent with following intuitive picture:
 {\it the bulk tachyon  counts the bulk degree of freedom 
  while the twisted tachyon counts the localized degree of 
  freedom.}
  So, the quantity measuring the localized quantity is the minimal 
  tachyon mass. We will evaluate it explicitly in later section.
  
\subsection{type II: no bulk tachyon}
Now we apply this result to type II case, where bulk tachyon 
is projected out by chiral GSO and the 
dominant contribution comes from  the lightest tachyon. 
As we have shown before, we can use the total partition function, since the dominant 
contribution is coming from twisted sector and that term is not 
cancelled out. 
Furthermore we get the same value of $c_{eff}$ whether we use
the full GSO projected partition function  or just its bosonic part.
The reason is very simple: when we conceptually write the total partition function   
as the  difference of bosonic and fermionic contributions,
\be
Z^{II}= Z^{II}_B- Z^{II}_F,
\ee
neither $Z^{II}_B$ nor  $Z^{II}_F$ should contain the bulk tachyon;
 it is simply not in the spectrum of the theory.
Therefore both quantities $Z^{II}_{tw}$,  $Z^{II}_{B, tw}$ 
give the same localized degree of freedom given by
\be
c_{eff}= 6|\alpha'M^2_{min}|, \label{ceff2}
\ee
which we will evaluate  shortly.

Now we meet interesting question for type II: Since there is no 
bulk tachyon, calculating the delocalized degree of freedom
\be
Z_{un}(\tau_2\to \infty) \sim g^{II}_{cl}q^0
\ee
shows that there is no bulk degree of freedom ($c_{eff}=0$). At first looking this 
looks puzzling. However, it is not so surprising by  taking the analogy of the minimal 
models, 
where we get $c<1$ minimal models by taking out the degrees of freedom from the Hilbert space of  
 the $c=1$ models. Along this line of thinking, spacetime supersymmetric theory is the case where all 
 the   degree of freedom is eliminated from 2 dimensional point of 
 view. From this point of view, even in the supersymmetric case, where 
$Z=0$,  $Z$  still correctly counts the central charge of the relevant 2d CFT.
This is also consistent with the c-theorem in the non-compact space:
the derivative of the central charge must be zero, hence the bulk degree of 
freedom should not change along the RG-flow, as pointed out by HKMM \cite{hkmm}.
It is 0 in type II and  12 in type 0.  These values do
  not change under the local tachyon condensation process. 
  Therefore the transition between the type 0 and type II is impossible from this point of view. 

\section{Minimal twisted tachyon mass as a c-function}
In the previous sections, we argued  that the c-function for the 
RG-flow in localized tachyon condensation is the minimal tachyon mass. 
We now explicitly evaluate the tachyon masses of  string theory on 
${\bf R}^{7,1} \times {\bf C}^1/{\bf Z}_N$.

\subsection{Type II orbifolds}
 Our starting point is the following 1-loop string vacuum amplitude of type II orbifold 
model:
\be
Z^{II} = \int_F \f{(d\tau)^2}{4\tau_2}(4\pi^2\al'\tau_2)^{-4} 
  Z^{II} (\tau), 
\ee
where \cite{Russo:1995ik,Takayanagi:2001jj,Lowe:1994ah} 
\ba 
Z^{II} (\tau) &=& \sum_{l,m=0}^{N-1} Z_{l,m}(\tau)\no
                 &=& \sum_{l,m=0}^{N-1}
\f{|\theta_3(\nu_{lm}|\tau)\theta_{3}(\tau)^3 
-\theta_2(\nu_{lm}|\tau)\theta_{2}(\tau)^3 
-\theta_4(\nu_{lm}|\tau)\theta_{4}(\tau)^3|^2} 
{4N|\eta(\tau)|^{18} |\theta_{1}(\nu_{lm}|\tau)|^2},
\ea
where $\nu^i_{l,m} = \frac{k_i}{N}(l-m\tau)$.
The dominant contribution comes  from the tachyons in the $\tau_2\to\infty $ limit. 
$Z_{tw}^{II}(\tau)$, $Z_{un}^{II}(\tau)$  are defined by
\be
 Z_{tw}^{II}(\tau) = \sum_{m=1}^{N-1}\sum_{l=0}^{N-1} Z_{l,m}(\tau) ,\;\;
  Z_{un}^{II}(\tau)=\sum_{l=0}^{N-1} Z_{l,0}(\tau). 
\ee
The low temperature limit of the partition function $Z_{tw}^{II}(\tau)$ 
can be easily evaluated to be 
\be
Z_{tw}^{II}(\tau_2\to \infty)=\sum_{m=1}^{N-1} q^{\half 
\alpha'M^2} \sim  \exp\left( \tau_2 |{\rm 
min}_{m=1}^{N-1} \pi \alpha'M^2|\right), 
\ee 
where 
\ba
\half \alpha' M^2 &=& -
\left\{ \frac{km}{N}\right\}, \quad {\rm if} 
\quad \left[\frac{km}{N}\right] \in 2{\mathbf Z}, \no
&=&  \left\{\frac{km}{N}\right\}-1, \quad  {\rm if} \quad 
\left[\frac{km}{N}\right] \in 2{\mathbf Z}+1,\no
\ea 
where we used the notation $\{x\}$ for the 
fractional part of $x$ and $[x]$ for the integer part of $x$ 

 The  minimal mass of the twisted tachyon can be easily seen to be 
\be
|{\rm min}_{m=1}^{N-1}  \alpha' 
M^2|=2\left(1-\frac{1}{N}\right).\label{d-angle}
\ee
Notice that it is independent of $k$ as long as $N,k$ are co-prime 
to each other. It is easy to prove above statement using the basic 
lemma of the number theory. From eq.(\ref{d}) we get 
$c_{\rm eff}=12\left(1-\frac{1}{N}\right).$
Notice that this value is proportional to the deficit angle, 
$\delta=2\pi(1-1/N)$ of the orbifold geometry. 

In a recent paper \cite{dabholkar}, Dabholkar and 
Vafa (DV) suggested that the tachyon potential should be given by the 
maximum value of the axial vector charge $Q_5=F_L+F_R$, i.e. 
$V(t, \bar{t})={\rm max}|Q^5|$.
For ${\bf C}/{\bf Z}_N$ case, one has $Q^5=1-1/N$. It is also well known 
that with N=2 worldsheet supersymmetry,  $Q^5$ plays the role of the effective central 
charge, $c_{eff}=12\times{\rm max}|Q^5|.$
Therefore, we can identify the potential as the minimal tachyon 
mass;
\be
V(t, \bar{t})= \half|{\rm min}\; \alpha ' M^2|.
\ee
Quite some time ago, Cecotti and Vafa\cite{Cecotti:1991me}, proposed that 
$Q^5$ is the generalized c-function  for d=2, N=2 SCFT.   They 
called this as algebraic c-function and partially proved the 
theorem near critical points. What  DV did is to adopt this c-function as the 
tachyon potential.  So, we now can say that the minimal tachyon 
mass is the c-function of the RG-flow. From now on we use the tachyon potential
 and c-function as the same words.  
 
In ${\bf C}^2/{\bf Z}_N$ model,
the tachyon spectrum again can be calculated by taking the 
$\tau_2\to \infty$ limit of the partition function: 
\be
Z^{II}(\tau)=\sum_{l,m=0}^{N-1} \f{|\theta_3(\nu^1_{l,m}|\tau)
\theta_3(\nu^2_{l,m}|\tau)\theta_3(\tau)^2
-\theta_2(\nu^1_{l,m}|\tau)\theta_2(\nu^2_{l,m}|\tau)\theta_2(\tau)^2 
-\theta_4(\nu^1_{l,m}|\tau)\theta_4(\nu^2_{l,m}|\tau)\theta_4(\tau)^2|^2}
{4N\ |\eta(\tau)|^{12}
|\theta_{1}(\nu^1_{l,m}|\tau)\theta_{1}
(\nu^2_{l,m}|\tau)|^{2}}. \label{PF22}
\ee 
The low temperature limit of  $Z_{tw}^{II}(\tau)$ is 
 \be
Z_{tw}^{II}(\tau_2\to \infty)=\sum_{m=1}^{N-1} \exp(-\pi \tau_2 \alpha'M^2),
\ee
where,
\ba
\half \alpha' M^2 &=& -\left|\left\{ \frac{k_1m}{N} \right\}-\left\{ \frac{k_2m}{N} 
\right\}\right|, \quad {\rm if} \quad \left[\frac{k_1m}{N}\right]+
\left[\frac{k_2m}{N}\right] \in 2{\mathbf Z}, \no
                    &=&  -\left|\left\{ \frac{k_1m}{N} \right\}+\left\{ \frac{k_2m}{N} 
\right\}-1\right|, \quad {\rm if} \quad \left[\frac{k_1m}{N}\right]+
\left[\frac{k_2m}{N}\right] \in 2{\mathbf Z}+1, \label{aps}   
\ea
here again  $\{x\}$ is the fractional part and $[x]$ is the integer part of $x$.
Notice that  the identification of the minimal 
mass as the  tachyon potential
\be 
V(N,k_1,k_2)=-{\rm min}_{m=1}^{N-1} \half \alpha' M^2(N,k_2,m),
\ee
still holds and it agrees with that of DV for $C^2/Z_N$ also.
Notice that 
the minimal mass  vanishes along the supersymmetric line $k_1=k_2$. 
The minimal mass of the twisted tachyon is not simple in terms of 
rational function of $k_1,k_2$ and $N$. For $k_2=2p+1, k_1=1$, however, we find 
\ba
  V(N,1,k_2)&=& (1 - 1/k_2)(1 - r/N), 
  \;\; if  \;\;  N \equiv r {\rm ~ mod ~} k_2, \; r = 1, 2,...,p \no
                                           &=& (1 - 1/k_2)(1 - r(1 + 1/p)/N), \;\;  if \;\; N \equiv -r {\rm ~ mod ~} k_2,
\ea
For $k_1 = 2$, some of the $V(N,2, k_2)$ can be  shown  to be 
\ba
V(N,2,4p) &=& 1 - (2p + 1)/N   , ~~~if~~~ N \equiv 1 {\rm ~ mod ~} 4p ,\no
V(N,2,4p + 2) &=& (1 - k_1/k_2)(1 - 1/N), ~~~if~~~N \equiv 1 {\rm ~ mod ~} 4p+2.
\ea
Other cases where both $k_1,k_2$ are bigger than 2,  are more complicated.  
Using these formula, it is easy to show that all of the process 
studied in APS are allowed by the potential. 
In figure 1, we give a contour plot of the potential in $k_1,k_2$  
plane. The darker the color, the lower the potential.

\begin{figure}[htbp1]
 \epsfxsize=80mm
\centerline{\epsfbox{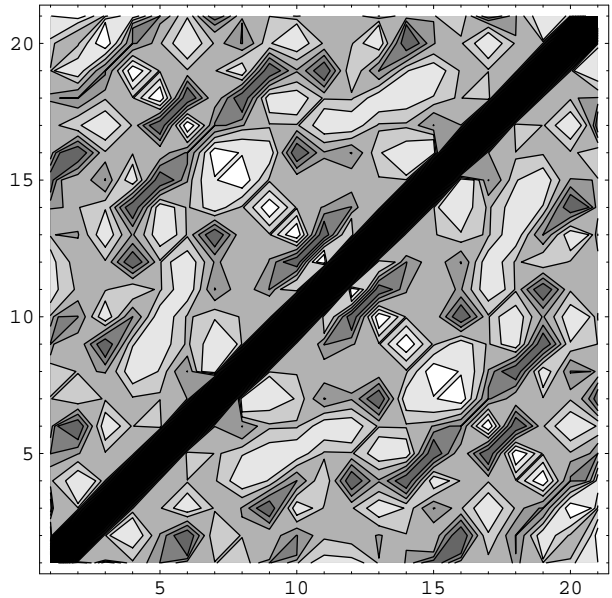}}
 \caption{Contour plot of type II's tachyon potential $V_{II}(N,k_1,k,2)= \half|{\rm min}\; \alpha ' M^2|$ in $(k_1,k_2)$ plane. Only
$k_1>0,k_2>0$ is shown for $N=23$. It has a valley along the line $k_1=k_2$, which guarantees the stability of the supersymmetric models.}
  \label{Fig1}
\end{figure}

The proposal that the minimal twisted tachyon mass as the c-function of 
the RG flow means the theory becomes less tachyonic along the phase 
transition. It is certainly  a generalized $c$-theorem: $c_{\rm eff}$ decreases along 
the RG-flow, which still need more rigorous proof.  
Since the $c_{\rm eff}$ represent the most dominant term  of the 
cosmological constant, decrease in tachyon mass is equivalent to 
the decrease in the cosmological constant. Therefore all these 
argument seems to support the speculation that in the absence of 
the tachyon potential coming from sphere amplitude,  torus 
partition function is the dominant contribution to the effective 
potential for the tachyon of closed superstring theory.    
 
\subsection{Type 0 } 
 
We now work out the c-function for the type 0 orbifolds by calculating  
 minimal mass of localized tachyons.
 First for $C^1/Z_N$ model, the partition function is 
\be
Z^0(\tau)= 
\sum_{l,m=0}^{N-1} 
 \f{|\theta_3(\nu_{lm}|\tau)\theta_{3}(\tau)^3|^2
+|\theta_2(\nu_{lm}|\tau)\theta_{2}(\tau)^3|^2
+|\theta_4(\nu_{lm}|\tau)\theta_{4}(\tau)^3|^2}
{2N|\eta(\tau)|^{18}
|\theta_{1}(\nu_{lm}|\tau)|^2},
\ee
with $k=odd$ and $(N,k)$ co-prime.
The low temperature limit of the twisted partition function $Z_{tw}^{0}(\tau)$ 
can be easily evaluated to be 
\be
Z_{tw}^{0}(\tau_2\to \infty)= \sum_{m=1}^{N-1} q^{\half 
\alpha'M^2 } 
\ee
where 
\be
\half \alpha' M^2 =  {\rm min}\left(- \left\{ \frac{km}{N}\right\}, \left\{\frac{km}{N}\right\}-1\right).  
\ee
So the localized degrees of freedom is counted by the dominant twisted tachyon 
term $q^{\half {\rm min}_{m=1}^{N-1}\alpha'M^2 }$, 
from which we  
 identify the  c-function for localized tachyon  for type 0;
\be
V(N,k)=\left|{\rm min}_{m=1}^{N-1} \half \alpha' M^2(N,k,m)\right| = 
{\rm max}_{m=1}^{N-1} {\rm max}\left( \left\{ \frac{km}{N}\right\}, 1-\left\{\frac{km}{N}\right\}\right).  
\ee
This looks slightly different from type II result. But in fact it gives identical 
result;
\be
V(N,k)= 1-\frac{1}{N}.
\ee

Notice that $g_{cl}q^{-1}$, which is coming from the bulk tachyon, is not present in $Z_{tw}$
in the low temperature limit.
It is contained in $Z_{un}$, counting the delocalized degrees of freedom.
So, we expect  that $g_{cl}$ can encode the information for the process 
of the localized tachyon condensation, 
due to the equivalence of the of the local and global geometry of the  orbifold
as mentioned before. This is discussed in separate publication \cite{ns1}.

Now we turn to type 0, $C^2/Z_N$ case.
The partition function is given by 
\be
Z^0(\tau) = \sum_{l,m=0}^{N-1}
\f{|\theta_3(\nu^1_{l,m}|\tau)
\theta_3(\nu^2_{l,m}|\tau)\theta_3(\tau)^2|^2
+|\theta_2(\nu^1_{l,m}|\tau)\theta_2(\nu^2_{l,m}|\tau)\theta_2(\tau)^2|^2 
+|\theta_4(\nu^1_{l,m}|\tau)\theta_4(\nu^2_{l,m}|\tau)\theta_4(\tau)^2|^2}
{2N|\eta(\tau)|^{12}
|\theta_{1}(\nu^1_{l,m}|\tau)\theta_{1}
(\nu^2_{l,m}|\tau)|^{2}}. \label{PF23}
\ee
with $k_1+k_2$ odd,  ${\rm gcd}(k_1,k_2,N)=1$. 

The low temperature limit of the partition function $Z_{tw}^{0}(\tau)$ is 
\be
Z_{tw}^{0}(\tau_2\to \infty)=  \sum_{m=1}^{N-1} q^{\half 
\alpha'M^2 }   
\ee
where 
\be
\half \alpha' M^2 =  {\rm min}\left(- \left |\left\{ \frac{k_1m}{N}\right\}-\left\{ \frac{k_2m}{N}\right\}\right|,
-\left| \left\{\frac{k_1m}{N}\right\}+\left\{\frac{k_2m}{N}\right\}-1\right|\right).  
\ee
Again the c-function is given by
\be
V(N,k_1,k_2)=\left|{\rm min}_{m=1}^{N-1} \half \alpha' M^2(N,k_1,k_2,m)\right| 
.
\ee
Instead of evaluating this in terms of rational functions,
we give a result of numerical analysis summarized in the contour plot in figure 2.
\begin{figure}[htbp2]
 \epsfxsize=80mm
\centerline{\epsfbox{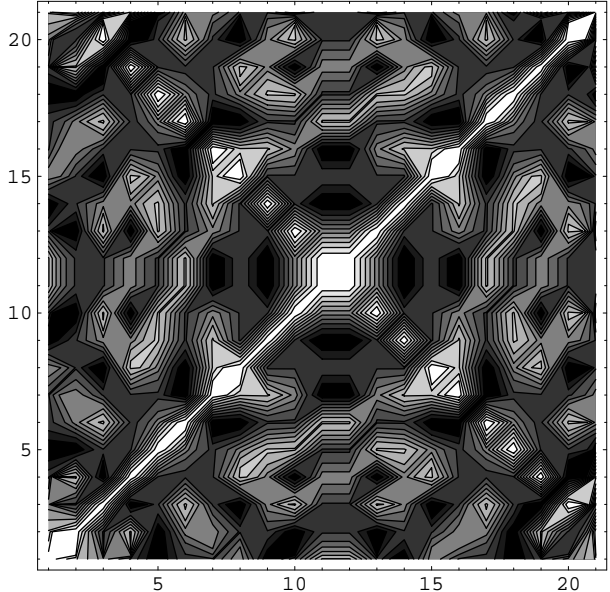}}
 \caption{Contour plot of  c-function, $V_{0}= \half|{\rm min}\; \alpha ' M^2|$, for type 0 $C^2/Z_N$ 
 model in $(k_1,k_2)$ plane. }
  \label{Fig2}
\end{figure}

\section{Summary and Discussion}
In this paper, we use cosmological constant to count the localized 
degree of freedom, and got following very intuitive results:
The bulk tachyon mass counts the bulk degree of freedom while 
the lowest twisted tachyon mass counts the localized degree of 
freedom. One can even refine the statement by saying that in each 
twisted sector, the lowest tachyon mass 
in each sector counts the degree of freedom in that sector. 
We also evaluated the lowest tachyon mass in the type 0 and type 
II orbifold models. We evaluated these minimal masses for 
$C^1/Z_N$, $C^2/Z_N$ both for type 0 and type II.  

Our results  suggest that the cosmological 
constant (1-loop string amplitude)  can play the role of tachyon 
potential.  According to it,  the  RG flow along the 
direction of more SUSY, less tachyonic, less singular, which is 
precisely what APS analysis has shown. This is a close analogue of 
open string result\cite{bgindep}, where effective action in 10 dimension is 
partition function at sphere. As shown by Tseytlin \cite{tseytlin}
there is no  tachyon potential coming from sphere. So naturally 
torus amplitude can play the leading tachyon potential.
In this approach it is necessary to related the localized nature 
of the tachyon potential and the divergence of the potential. 

Finally, we discuss future problems. 
Obviously, the most 
demanding problem is actually  to prove the generalized c-theorem: 
tachyon becomes less tachyonic along the RG-flow along the line of original Zamolochikov's\cite{Zamolodchikov:gt}  for the localized 
degree of freedom. It should be clarified explicitly why the $c_{eff}$ for 
the localized degree of freedom can be changed while the central charge for the 
bulk degree of freedom can not. 
In this paper, we mainly discussed the localized degree of freedom and got a c-function given by the 
minimal tachyon mass. However, the bulk degrees of 
freedom also encodes some information on the precess of localized 
condensation since local geometry around the fixed points completely fixes the global geometry
for the orbifolds. In fact, by considering the delocalized degree of freedom, 
we arrive at the quantity $g_{cl}$ for type 0 appeared in \cite{hkmm} and  
$ g_{cl}^{II}$ discussed in \cite{ns1} as a c-function. 
Then the detailed relation between these different criteria ( c-functions v.s g-functions) 
are desired. We will come back to this issue in future 
publication\cite{leesin}. 
 
{\bf Acknowledgements}  \\
The author wish to thank David Kutasov for critical comments on 
his previous work, to Shiraz Minwalla and Soonkeon Nam for communications on 
c-theorem,  and to Joe Polchinski for his interests and helpful remarks on the issue. 
This work is supported by KOSEF 1999-2-112-003-5.

\end{document}